\documentclass[prc,twocolumn,preprintnumbers,amsmath,amssymb,superscriptaddress,floatfix,nofootinbib]{revtex4}
\usepackage{graphicx}
\usepackage{amsmath}
\usepackage{amsfonts}
\usepackage{slashed}
\usepackage{amssymb}
\usepackage{hyperref}
\usepackage{slashed}

\newcommand{\smsm}[1]{\scriptscriptstyle{\scriptscriptstyle{#1}}}

\def\CM{{\cal M}}

\begin{document}

\title{Roles of $\Lambda$ resonances in the $K^- p \to \gamma \Sigma$ reaction}
\author{Yi Pan}
\author{Bo-Chao Liu}
\email{liubc@xjtu.edu.cn}

\address{Ministry of Education Key Laboratory for Nonequilibrium Synthesis
and Modulation of Condensed Matter, Shaanxi Province Key Laboratory of Quantum Information and
Quantum Optoelectronic Devices, School of Physics, Xi’an Jiaotong University, Xi’an 710049, China}

\begin{abstract}
We investigate the $K^- p \to \gamma \Sigma$ reaction using an effective Lagrangian approach within an isobar model framework. The model includes contributions from $s$-channel hyperon and hyperon resonance, $t$-channel $K$ and $K^*$, $u$-channel proton and $\Delta(1232)$ exchanges, and a phenomenological contact term. Our analysis focuses on the roles of various $\Lambda$ resonances. The results indicate that the $\Lambda(1600)$ resonance is crucial for describing the experimental data at lower energies. At higher energies, the inclusion of either the $\Lambda(1670)$ or the $\Lambda(1690)$ resonance significantly improves the agreement with data. However, the current data are insufficient to distinguish between these two scenarios. We suggest that future measurements at higher beam energies and the measurement of the $\Sigma$ polarization are needed to identify the roles of the $\Lambda(1670)$ and $\Lambda(1690)$ resonances.
\end{abstract}
\maketitle
\section{INTRODUCTION}
\label{sec:introduction}
The study of interactions between kaon beams and nucleon targets, such as the $K^- p$ reaction, provides an indispensable tool for exploring the spectrum of hyperon resonances ($Y^*$) and the intricate dynamics of the strong force within the strangeness sector. While pion-nucleon ($\pi N$) scattering has historically laid the groundwork for baryon spectroscopy, initial states with strangeness ($S=-1$) offer a unique window into excited hyperons. Such investigations are crucial for testing the predictions of SU(3) flavor symmetry and deepening our understanding of non-perturbative hadronic physics. Despite their importance, progress in this field has been significantly hampered by the quality of available experimental data. With the exception of recent results from the Crystal Ball Collaboration\cite{EXData1_Prakhov,EXData2_TDS}, the majority of the world's $K^- p$ scattering data were collected several decades ago and consequently suffer from low statistics and large uncertainties. Thus, new high-precision data are urgently needed to advance our knowledge of hyperon resonances and the mechanisms governing their production. This data scarcity is expected to be addressed by planned experiments with new kaon beams at JLab\cite{KLF:2020gai} and J-PARC\cite{Nagae:2008zz,Ohnishi:2019cif}.

In this work, we investigate the \( K^- p \to \gamma \Sigma^0 \) reaction, motivated by recent measurements from the Crystal Ball Collaboration~\cite{EXData1_Prakhov}.  
Historically, kaon-induced radiative reactions were studied experimentally in the 1960s--1980s~\cite{Mast:1968ltv,Colas:1975ck,Bertini}, but the available data were sparse and of limited precision. This process is related by crossing symmetry to the extensively studied photoproduction channel \( \gamma p \to K^+ \Sigma^0 \), providing a complementary probe of the underlying reaction dynamics~\cite{Ji:1988zza,Williams:1990hh,Williams:1991tw,David:1995pi}.  
The \( \gamma p \to K^+ \Sigma^0 \) channel has been the subject of intense theoretical and experimental investigation for decades. While early theoretical models\cite{ref:theory_early,ref:theory_early2,ref:theory_early3,ref:theory_early4} were constrained by limited data, the landscape was transformed by high-precision measurements from the CLAS and LEPS collaborations. These data stimulated a new wave of theoretical analyses employing a variety of frameworks, including isobar models\cite{ref:isobar,ref:isobar2,ref:isobar3,ref:isobar4,couple_SNK,ref:isobar5,ref:isobar6}, dynamical coupled-channel approaches\cite{ref:dcc,ref:dcc2,ref:dcc3,ref:dcc4}, and the Regge-plus-resonance (RPR) formalism\cite{ref:rpr,ref:rpr1,ref:rpr2}. Among these, isobar models have proven particularly effective in describing photoproduction data. However, different implementations, while often achieving comparable agreement with experiment, frequently rely on disparate assumptions regarding the underlying dynamics. Key differences arise in the choice of contributing resonances, coupling constants, hadronic form factors, and schemes for enforcing gauge invariance. These discrepancies highlight significant model-dependent ambiguities in our theoretical understanding. Applying and testing these models in the crossing channel, $K^- p \to \gamma \Sigma^0$, provides a powerful consistency check and is helpful for discriminating between competing theoretical descriptions and ultimately refining our knowledge of the reaction mechanisms.

As mentioned above, the most recent experimental data for $K^- p \to \gamma \Sigma^0$ were published by the Crystal Ball collaboration\cite{EXData1_Prakhov}. Their accompanying analysis employed an RPR model that considered only one hyperon resonance---either $\Lambda(1600)$, $\Lambda(1670)$, or $\Lambda(1690)$---at a time, alongside a Reggeized background. This simplified approach failed to provide a satisfactory description of the data, suggesting that the interplay between resonances and the treatment of the non-resonant background are more complex than assumed. This motivates a comprehensive re-examination of the reaction dynamics, incorporating more resonances and a robust treatment of the background.

A central challenge in modeling such reactions is ensuring gauge invariance, which is often violated when hadronic form factors are introduced. To address this fundamental requirement, we employ a well-established method for restoring gauge invariance\cite{gague_restore,gague_restore1,gague_restore2,couple_SNK}, which has been successfully applied in previous studies of the photoproduction channels, e.g. $\gamma N \to K\Sigma$. The primary objective of this study is twofold. First, we develop a more sophisticated, manifestly gauge-invariant model to assess if it improves the description of existing data. Second, we seek to investigate the collective role of multiple hyperon resonances, specifically probing the combined contributions of the $\Lambda(1600)$ state together with either the $\Lambda(1670)$ or the $\Lambda(1690)$. Through this analysis, we aim to elucidate the importance of resonance interference effects and develop a more physically complete and consistent picture of the $K^- p \to \gamma \Sigma^0$ reaction.

This paper is organized as follows. In Sec.~\ref{sec:MODEL}, we present the theoretical formalism for the reaction $K^- p \to \gamma \Sigma$, including the method for restoring gauge invariance and the key components of our model. In Sec.~\ref{sec:discussion}, we present and discuss the numerical results. Finally, we conclude with a summary in Sec.~\ref{sec:summary}.

\section{MODEL AND INGREDIENTS}\label{sec:MODEL}
In this work, based on the effective Lagrangian approach and isobar model, we construct a dynamical model with background terms including $t$-channel $K$ and $K^*$ meson exchanges, $u$-channel proton and $\Delta^+$ baryon exchanges, $s$-channel $\Lambda$ and $\Sigma$ hyperon exchanges, as well as contact interactions. Current knowledge for the electromagnetic coupling of hyperon resonances is still very limited. Here we follow Ref.\cite{EXData1_Prakhov} and incorporate the contributions from the hyperon resonances $\Lambda(1600)(1/2^+)$, $\Lambda(1670)(1/2^-)$, and $\Lambda(1690)(3/2^-)$ in the $s$-channel, since these resonances are predicted to have relatively large radiative decay branching ratios in quark model\cite{branching,branching1}.
\begin{figure}[htbp]
	\begin{center}
		\includegraphics[scale=0.7]{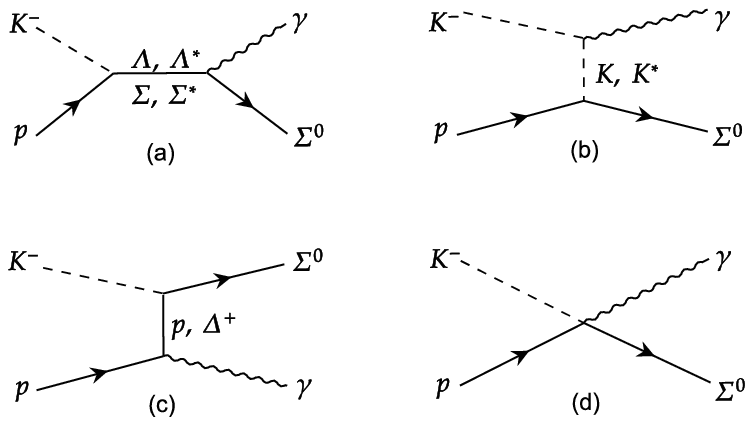} \vspace{-0.2cm}
		\caption{Feynman diagrams for the $K^- p \rightarrow \gamma \Sigma$ reaction with $\Lambda^{*}$ denoting the $\Lambda(1600)$, $\Lambda(1670)$ or $\Lambda(1690)$.} \label{feynfig}
	\end{center}
\end{figure}

We adopt the interaction Lagrangian densities from Refs.\cite{couple_SNK,couple_LNK_1_SNKstar_1,Kappa_SNKstar_GLS_formfactor_1} to describe the couplings at the relevant vertices involving hyperons and hyperon resonances.
\begin{align}
\mathcal{L}_{\Sigma\Sigma\gamma} &= -e\bar{\Sigma}\left[\left(\hat{e}\gamma^{\mu} - \frac{\hat{k}_{\Sigma}}{2M_{\Sigma}}\sigma^{\mu\nu}\partial_{\nu}\right)A_{\mu}\right]\Sigma,\\
\mathcal{L}_{\gamma\Lambda\Sigma} &= \frac{e\kappa_{\Sigma\Lambda}}{2 M_{\Lambda}} \bar{\Lambda} \sigma^{\mu\nu} \partial_\nu A_\mu \Sigma + \text{H.c.}, \\
\mathcal{L}_{R\Sigma\gamma}^{1/2^\pm} &= e \frac{g_{R\Sigma\gamma}}{2 M_\Sigma} \bar{R} \Gamma^{(\mp)} \sigma_{\mu\nu} (\partial^{\nu} A^{\mu}) \Sigma + \text{H.c.}, \\
\mathcal{L}_{R\Sigma\gamma}^{3/2^\pm} &= -i e \frac{g_{R\Sigma\gamma}^{(1)}}{2 M_{\Sigma}} \bar{R}_{\mu} \gamma_{\nu} \Gamma^{(\pm)} F^{\mu\nu} \Sigma \nonumber \\
&\quad + e \frac{g_{R\Sigma\gamma}^{(2)}}{(2 M_{\Sigma})^{2}} \bar{R}_{\mu} \Gamma^{(\pm)} (\partial_{\nu} F^{\mu\nu}) \Sigma + \text{H.c.}, \\
\mathcal{L}_{\Lambda NK} &= \frac{g_{\Lambda NK}}{2M_N} \bar{\Lambda} \gamma_5 \gamma^{\mu} (\partial_\mu K) N + \text{H.c.}, \\
\mathcal{L}_{\Sigma NK} &= \frac{g_{\Sigma NK}}{2M_N} \bar{\Sigma} \cdot \vec{\tau} \gamma_5 \gamma^{\mu} (\partial_\mu \vec{K}) N + \text{H.c.}, \\
\mathcal{L}_{RNK}^{1/2^\pm} &= \frac{g_{RNK}}{M_N+M_R}\bar{N} \Gamma^{(\pm)}\gamma_5 \gamma^{\mu} (\partial_\mu K) R + \text{H.c.}, \\
\mathcal{L}_{RNK}^{3/2^\pm} &= \pm \frac{g_{RNK}}{M_K} \bar{N} \Gamma^{(\mp)} (\partial^{\alpha} K) R_{\alpha} + \text{H.c.},
\end{align}
where $R$ designates the $\Lambda(1600)$, $\Lambda(1670)$, or $\Lambda(1690)$ resonance, and the superscript on $\mathcal{L}_{R\Sigma\gamma}$ and $\mathcal{L}_{RNK}$ denotes the $J^P$ quantum numbers of the resonance $R$. The parity operators $\Gamma^{(+)}$ and $\Gamma^{(-)}$ are defined as $\gamma_5$ and $1$, respectively. In these Lagrangians, \( e \) denotes the elementary charge, \( \hat{\kappa}_\Sigma = \frac{1 + \hat{e}}{2}\,\kappa_{\Sigma^+} + \frac{1 - \hat{e}}{2}\,\kappa_{\Sigma^-} \), where the anomalous magnetic moments are \( \kappa_{\Sigma^+} = 1.458 \) and \( \kappa_{\Sigma^-} = -0.16 \). Additionally, \( \kappa_{\Sigma\Lambda} = 1.61 \) is the anomalous magnetic moment associated with the \( \Sigma^0 \to \Lambda\gamma \) transition~\cite{ParticleDataGroup:2024cfk}. The coupling constants $g_{\Lambda NK}$ and $g_{\Sigma NK}$ are taken to be $-14$\cite{couple_LNK1,couple_LNK_2} and $2.69$\cite{couple_SNK}, respectively. The coupling constants $g_{RNK}$ can be obtained from the partial decay widths provided by the Particle Data Group (PDG)\cite{ParticleDataGroup:2024cfk}. However, the couplings for the $R\Sigma\gamma$ vertices are not well-determined. Their values will be fit to the experimental data as discussed in Sec.~\ref{sec:discussion}. 

For the $t$-channel meson exchange contributions, we consider the exchange of $K$ and $K^*$ mesons. The relevant interaction Lagrangian densities are taken from Ref.\cite{couple_SNK,Lagrangian_GamaKKstar}:
\begin{align}
\mathcal{L}_{\gamma KK} &= ie \left[ K^{+} (\partial_\mu K^{-}) - K^{-} (\partial_\mu K^{+}) \right] A^\mu, \\
\mathcal{L}_{\gamma K K^*} &= e \frac{g_{\gamma K K^*}}{M_K} \varepsilon^{\mu\nu\rho\sigma} (\partial_\mu A_\nu) (\partial_\rho K_\sigma^*) K, \\
\mathcal{L}_{\Sigma N K^*} &= -g_{\Sigma N K^*} \bar{\Sigma} \left[ \left( \gamma^{\mu} - \frac{\kappa_{\Sigma N K^*}}{2 M_N} \sigma^{\mu\nu} \partial_{\nu} \right) K_{\mu}^* \right] N \nonumber \\
&\quad + \text{H.c.},
\end{align}
where $K_{\mu}^{*}$ represents the $K^{*}$ field. For the electromagnetic coupling $g_{\gamma K K^*}$, the coupling $g_{\Sigma NK^*}$ and the tensor coupling $\kappa_{\Sigma N K^*}$, we adopt the values from Refs.\cite{couple_SNK,couple_SNKs}.

For the $u$-channel diagrams, we consider the exchange of a proton and a $\Delta^{+}$ baryon. The required interaction Lagrangians are given as follows:
\begin{align}
\mathcal{L}_{\gamma NN} &= -e \bar{N} \left[ \left( \gamma^\mu - \frac{\kappa_N}{2 M_N} \sigma^{\mu\nu} \partial_\nu \right) A_\mu \right] N, \\
\mathcal{L}_{\Delta \Sigma K} &= \frac{g_{\Delta \Sigma K}}{M_K} \bar{\Sigma} (\partial^{\mu} K) \Delta_\mu + \text{H.c.}, \\
\mathcal{L}_{\Delta N \gamma} &= -i e \frac{g_{\Delta N \gamma}^{(1)}}{2 M_{N}} \bar{\Delta}_{\mu} \gamma_{\nu} \gamma_{5} F^{\mu\nu} N \nonumber \\
&\quad + e \frac{g_{\Delta N \gamma}^{(2)}}{(2 M_{N})^{2}} \bar{\Delta}_{\mu} \gamma_{5} (\partial_{\nu} F^{\mu\nu}) N + \text{H.c.},
\end{align}
where $\kappa_p = 1.793$ is the anomalous magnetic moment for the proton\cite{ParticleDataGroup:2024cfk}. The coupling constants $g_{\Delta \Sigma K}$, $g_{\Delta N \gamma}^{(1)}$, and $g_{\Delta N \gamma}^{(2)}$ are taken from Ref.\cite{couple_SNK}.

Finally, a contact term is included to ensure the overall gauge invariance of the amplitude. The corresponding interaction Lagrangian is defined as
\begin{align}
\mathcal{L}_{\gamma \Sigma N K} = i e \frac{g_{\Sigma N K}}{2 M_N} \bar{\Sigma} \gamma^{\mu} \gamma_5 A_{\mu} K N + \text{H.c.}.
\label{eq:L_contact}
\end{align}

It is well known that introducing form factors can violate the gauge invariance of the Born-term amplitude\cite{gague-violating_1,gague-violating_gague_restore_1}.  In this work, we follow the procedure from Refs.\cite{gague_restore,gague_restore1,gague_restore2,couple_SNK,Pan:2025fzg} to restore gauge invariance and introduce an additional term of the form:
\begin{equation}
    \mathcal{M}^{\mu}_{\text{add}} = - \frac{ig_{\Sigma N K}}{2 M_N} \slashed{p}_K C^{\mu}.
\end{equation}
The auxiliary current $C^{\mu}$ is given by:
\begin{equation*}
    C^{\mu} = -e_{K^-} \frac{f_{K}-\hat{F}}{t-p_{K}^{2}}(2 p_{K}-k)^{\mu} - e_{p} \frac{F_{p}-\hat{F}}{u-p_{p}^{2}}(2 p_{p}-k)^{\mu},
\end{equation*}
where $\hat{F} = 1 - \hat{h}(1 - F_{p})(1 - f_{K})$. In principle, $\hat{h}$ is an arbitrary function of Mandelstam invariants that approaches unity in the high-energy limit. For simplicity, it is commonly set to $\hat{h}=1$, and we adopt this choice in the present work.

The propagators for intermediate states with spin $J$, denoted as $S_J$, are taken as
\begin{align}
    S_{0}(q) &= \frac{i}{q^2-m^2}, \label{pp1} \\
    S_{1}^{\mu\nu}(q) &= -\frac{i(g^{\mu\nu}-q^\mu q^\nu/m^2)}{q^2-m^2}, \label{pp2} \\
    S_{1/2}(q) &= \frac{i(\slashed{q}+m)}{q^2-m^2+im\Gamma}, \label{pp3} \\
    S_{3/2}^{\mu\nu}(q) &= \frac{i(\slashed{q}+m)}{q^2-m^2+im\Gamma} \left[ -g^{\mu\nu} + \frac{1}{3}\gamma^\mu\gamma^\nu \right. \nonumber \\
    &\quad \left. + \frac{1}{3m}(\gamma^\mu q^\nu - \gamma^\nu q^\mu) + \frac{2}{3m^2}q^\mu q^\nu \right],
\end{align}
where $q$, $m$, and $\Gamma$ are the four-momentum, mass, and decay width of the intermediate particle, respectively.

To take into account the internal structure of hadrons and off-shell effects, phenomenological form factors are introduced. For baryon and baryon resonance exchanges, we use the following form factor\cite{couple_SNK,ref:isobar5}:
\begin{equation}
    F_B(q^{2}) = \left( \frac{\Lambda_{B}^{4}}{\Lambda_{B}^{4}+(q^{2}-m^{2})^{2}} \right)^2,
    \label{FF1}
\end{equation}
where $q$ and $m$ are the four-momentum and mass of the exchanged baryon, and $\Lambda_{B}$ is the corresponding cutoff parameter. For meson exchanges, we use the form factor from Refs.\cite{couple_SNK,ref:isobar5}:
\begin{equation}
    f_M(q^2) = \left( \frac{\Lambda_M^2-m^2}{\Lambda_M^2-q^2} \right)^2,
    \label{FF2}
\end{equation}
where $\Lambda_{M}$ is the cutoff parameter for the exchanged meson. The values of the parameters adopted in this work are collected in Tab.~\ref{parameters}. 

It is worth noting that the cutoff parameters $\Lambda_{N(\Delta)}$, $\Lambda_{K(K^*)}$, and $\Lambda_{\Sigma(\Lambda)}$ are taken from Ref.~\cite{ref:isobar5}, which studied $\gamma n \to K^0 \Sigma^0$ and $\gamma n \to K^+ \Sigma^-$ using an identical effective Lagrangian approach and the same gauge-invariance prescription. Given the shared interaction vertices (e.g., $NK^*K$, $N\Delta$, $\Sigma N\Lambda$), these cutoffs provide a well-motivated reference for our model. 

For $t$-channel meson exchanges, the magnitude of the momentum transfer in the crossed reactions is similar, supporting the use of the same cutoffs. In contrast, for baryon exchanges in the $s$- and $u$-channels, the kinematical regions differ between crossed processes, which could in principle lead to different optimal cutoff values. Nevertheless, for simplicity and to minimize the number of free parameters, we adopt these values in our calculations. 

Furthermore, we also fix the cutoff $\Lambda_R$ at 1.3~GeV, within the range suggested in Ref.~\cite{formfactorL13}. We have explicitly verified that the specific choices of all these fixed cutoff parameters—within reasonable ranges—do not significantly affect the main conclusions of this work.

       \begin{table}[htbp]
        \caption{Values of parameters adopted in this work.}
        \begin{tabular}{cccc}
        \hline\hline
Parameters                      & Values                    &  Parameters                   & Values\\
\hline
$g_{\Sigma N K}$                &$ 2.69$\cite{couple_SNK,Delta}                                  &$g_{\Delta N {\gamma}}^{(1)}$        &$-4.18$    \cite{couple_SNK,Delta}\\
$g_{\Sigma N K^*}$             &$ -4.26$\cite{couple_SNK,couple_SNKs}                  &$g_{\Delta N {\gamma}}^{(2)}$     &$4.327$    \cite{couple_SNK,Delta}\\
$g_{\gamma K K^*}$              &$ 0.413$\cite{couple_SNK,couple_SNKs}                                          &$g_{\Delta \Sigma K}$            &$7.89$ \cite{couple_SNK}\\
$g_{\Lambda N K}$               &$ -14$\cite{couple_LNK_2,couple_LNK1}                &$\kappa_{\Sigma NK^*}$                 &$ -2.33$\cite{couple_SNK,couple_SNKs}\\
$\kappa_{p}$                   &$ 1.793$ \cite{ParticleDataGroup:2024cfk}                       &$\Lambda_{K(K^*)}$       &$ 0.589 $\cite{ref:isobar5}\\
$\kappa_{\Sigma\Lambda}$              &$ 1.61$\cite{ParticleDataGroup:2024cfk}                         &$\Lambda_{\Sigma(\Lambda)}$         &$ 1.218 $\cite{ref:isobar5}           \\
$\Lambda_{N(\Delta)}$                   &$ 0.984 $\cite{ref:isobar5}                                                 &$\Lambda_{R}$       &$ 1.3 $\cite{formfactorL13}\\
\hline \hline
        \end{tabular}
        \label{parameters}
    \end{table}

The general amplitude for the $K^- p \to \gamma \Sigma$ reaction can be written as
\begin{equation}
    M = \bar{u}_{\Sigma} \mathcal{M}^{\mu} \epsilon_{\mu}^* u_{N},
\end{equation}
where $u_{N}$ and $\bar{u}_{\Sigma}$ are the Dirac spinors for the initial proton and the final $\Sigma$ hyperon, respectively, and $\epsilon_{\mu}^*$ denotes the polarization vector of the outgoing photon. With the ingredients defined above, the individual scattering amplitudes are obtained as follows:
\begin{eqnarray}
\CM^{\mu,1}_{\frac{3}{2}^-}&=&i\frac{eg^{(1)}_{R\Sigma\gamma}g_{RNK}}{2M_\Sigma M_K} (k_\nu \gamma^\mu S^{\nu\rho}_{\frac{3}{2}}-\not kS^{\mu\rho}_{\frac{3}{2}}) p^K_\rho\gamma_5 F_R, \nonumber\\
\CM^{\mu,2}_{\frac{3}{2}^-}&=&i\frac{eg^{(2)}_{R\Sigma\gamma}}{(2M_\Sigma)^2}\frac{g_{RNK}}{M_K} (k_\nu p_\Sigma^\mu S^{\nu\rho}_{\frac{3}{2}}-k\cdot p_\Sigma S^{\mu\rho}_{\frac{3}{2}}) p^K_\rho \gamma_5 F_R,\nonumber\\
M^\mu_{\frac{1}{2}^+}&=&i\frac{eg_{R\Sigma\gamma}}{2M_\Sigma}\frac{g_{RNK}}{M_R+ M_N} k_\nu  \sigma^{\mu\nu} S_{\frac{1}{2}} \not p_K\gamma_5 F_R, \nonumber\\
M^\mu_{\frac{1}{2}^-}&=&i\frac{eg_{R\Sigma\gamma}}{2M_\Sigma}\frac{g_{RNK}}{M_R+ M_N} k_\nu  \sigma^{\mu\nu}\gamma_5 S_{\frac{1}{2}} \not p_K F_R, \nonumber\\
\CM_N^\mu&=&-i\frac{eg_{\Sigma NK}}{2M_N} \not p_K\gamma_5S_{\frac{1}{2}}(\gamma^\mu-\frac{i\kappa_N}{2M_N}\sigma^{\mu\sigma}k_\sigma) F_p,\nonumber\\
\CM_{\Delta}^\mu&=&\frac{eg^{(1)}_{\Delta N \gamma}g_{\Delta \Sigma K}}{2M_N M_K} p^K_\rho(k_\nu S^{\nu\rho}_{\frac{3}{2}}\gamma^\mu -S^{\mu\rho}_{\frac{3}{2}} \not k) \gamma_5 F_\Delta+ \nonumber\\ 
& &\frac{eg^{(2)}_{\Delta N \gamma}g_{\Delta \Sigma K}}{({2M_N})^2M_K} p^K_\rho(k_\nu p_N^\mu S^{\nu\rho}_{\frac{3}{2}}-k\cdot p_N S^{\mu\rho}_{\frac{3}{2}})  \nonumber \gamma_5 F_\Delta, 
\end{eqnarray}
\begin{eqnarray}
\CM_K^\mu &=&-\frac{ieg_{\Sigma NK}}{M_N}p_K^\mu S_0(\not k-\not p_K)\gamma_5 f_K, \nonumber\\
\CM_{K^*}^\mu&=&-\frac{eg_{\gamma KK^*}g_{\Sigma NK^*}}{M_K}\varepsilon^{\mu \nu \rho \sigma}k_\nu (k-p_K)_\rho S^1_{\sigma\lambda} (\gamma^\lambda\nonumber\\&&-\frac{i\kappa_{K*}}{2M_N}\sigma^{\lambda\delta}(k-p_K)_\delta) f_{K^*},\nonumber\\
\CM_\Lambda^\mu&=&-\frac{eg_{\Lambda NK}\kappa_{\Sigma\Lambda}}{4M_NM_\Lambda}\sigma^{\mu\nu}k_\nu S_{\frac{1}{2}}\not p_K \gamma_5 F_\Lambda,\nonumber\\
\CM_\Sigma^\mu
&=&-\frac{eg_{\Sigma NK}\kappa_\Sigma}{4M_NM_\Sigma}\sigma^{\mu\nu}k_\nu S_{\frac{1}{2}}\not p_K \gamma_5 F_\Sigma,\nonumber\\
\CM_c^\mu&=&-i\frac{eg_{\Sigma NK}}{2M_N}\gamma^\mu\gamma_5 f_C.\nonumber
\end{eqnarray}
Here, the superscripts 1 and 2 for $\mathcal{M}^\mu_{\frac{3}{2}^-}$ denote the amplitudes corresponding to the Lagrangians $\mathcal{L}^{3/2^-}_{R\Sigma\gamma,(1)}$ and $\mathcal{L}^{3/2^-}_{R\Sigma\gamma,(2)}$, respectively.

The unpolarized differential cross section for the reaction $K^- p \to \gamma \Sigma$ in the center-of-mass (c.m.) frame is given by
\begin{equation}
\frac{d\sigma}{d\Omega} = \frac{1}{2} \frac{M_p M_\Sigma}{32\pi^2 s} \frac{|\mathbf{k}|}{|\mathbf{p}_{K}|} \sum_{\lambda, s_p, s_\Sigma} |\mathcal{M}_{fi}|^2,
\end{equation}
where $\mathcal{M}_{fi}$ is the total transition amplitude, and $|\mathbf{k}|$ and $|\mathbf{p}_{K}|$ are the magnitudes of the three-momenta of the final photon and the initial $K^-$ meson in the c.m. frame, respectively. The quantity $s = (p_K + p_p)^2$ is the Mandelstam variable, representing the square of the total c.m. energy, with $p_K$ and $p_p$ being the four-momenta of the $K^-$ and the proton. The summation runs over the final photon helicity $\lambda$ and the spin projections of the initial proton ($s_p$) and final $\Sigma$ hyperon ($s_\Sigma$). The factor of $1/2$ accounts for the average over the two possible spin states of the unpolarized initial proton.

Furthermore, the $\Sigma$ polarization asymmetry, $P_\Sigma$, can be calculated as
\begin{equation}
P_\Sigma = \frac{d\sigma(\uparrow) - d\sigma(\downarrow)}{d\sigma(\uparrow) + d\sigma(\downarrow)},
\end{equation}
where $d\sigma(\uparrow)$ and $d\sigma(\downarrow)$ are the differential cross sections for producing a $\Sigma$ hyperon with spin projection $+1/2$ and $-1/2$, respectively. The polarization axis is defined to be normal to the scattering plane, i.e., along the direction of $\mathbf{p}_K \times \mathbf{k}$ in the c.m. frame. The denominator, $d\sigma(\uparrow) + d\sigma(\downarrow)$, corresponds to the unpolarized differential cross section given above.

\section{RESULTS AND DISCUSSIONS}\label{sec:discussion}

In this section, we present and discuss our calculated results in comparison with the recent experimental data. The data for the $K^- p \to \gamma\Sigma$ reaction are taken from Ref.\cite{EXData1_Prakhov}, which reported angular distributions at eight incident $K^-$ momenta between 514 and 750~MeV. This corresponds to a total center-of-mass (c.m.) energy, $W$, spanning the range of 1.588 to 1.676~GeV. In the energy region under study, a number of hyperon resonances exist that may play important roles. However, as discussed in the preceding section, due to the limited information on their radiative decays, we only consider three specific resonances predicted to have relatively large radiative decay branching ratios in quark models\cite{branching,branching1}: the $\Lambda(1600)$, $\Lambda(1670)$, and $\Lambda(1690)$.
\begin{figure}[htbp]
	\begin{center}
		\hspace*{-5mm}
		\includegraphics[scale=0.4]{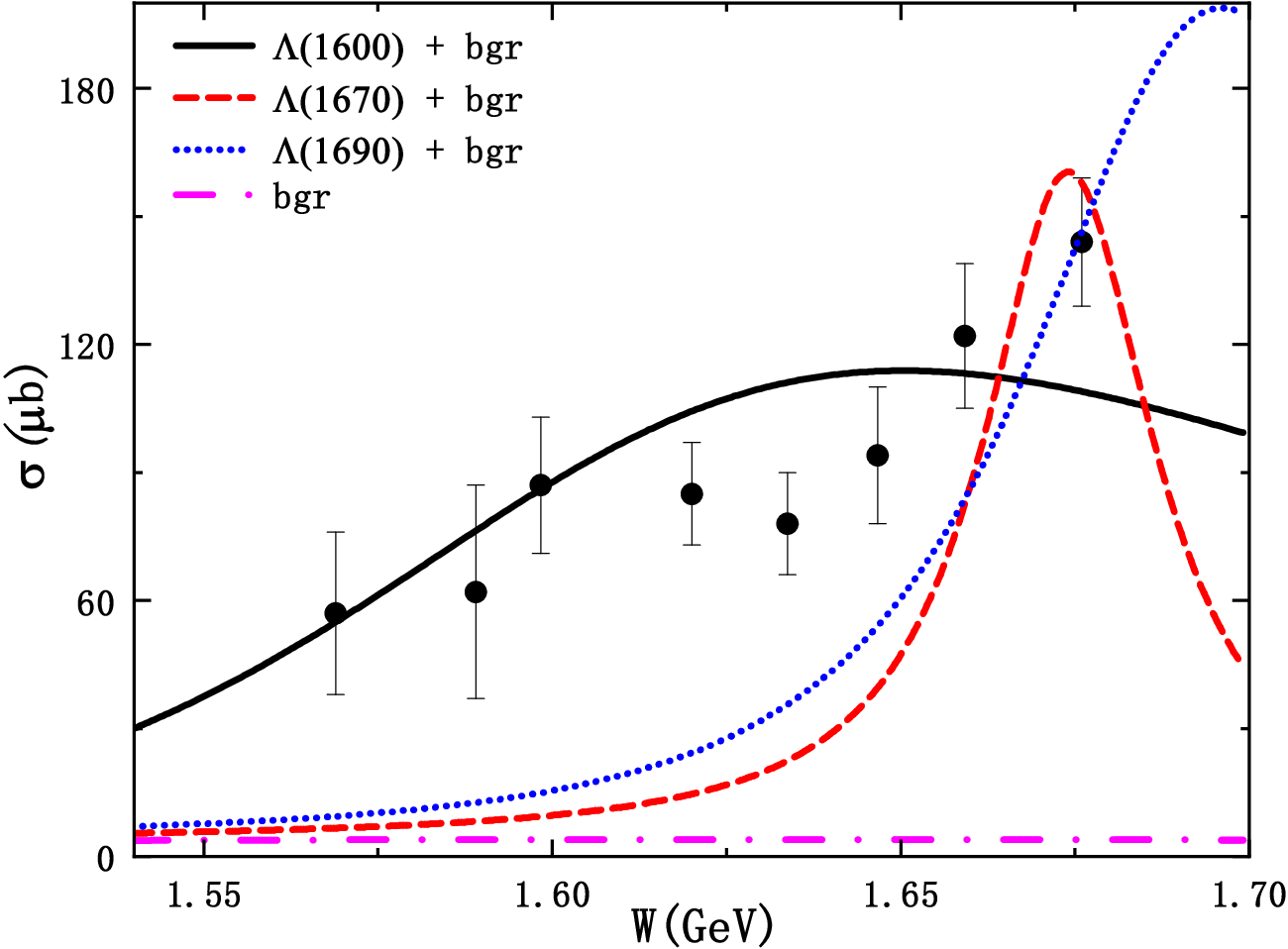} 
		\caption{Fitted total cross sections for the $K^- p \rightarrow \gamma \Sigma$ reaction considering the background and one of the three $\Lambda$ resonances: $\Lambda(1600)$(solid line), $\Lambda(1670)$(dashed line) and $\Lambda(1690)$(dotted line). The individual contribution of background terms is shown by the dash-dotted line.}
		\label{fig:xsec1R}
	\end{center}
\end{figure}
\begin{figure*}[htbp]
	\begin{center}
		\hspace*{-25mm}		\includegraphics[scale=0.8]{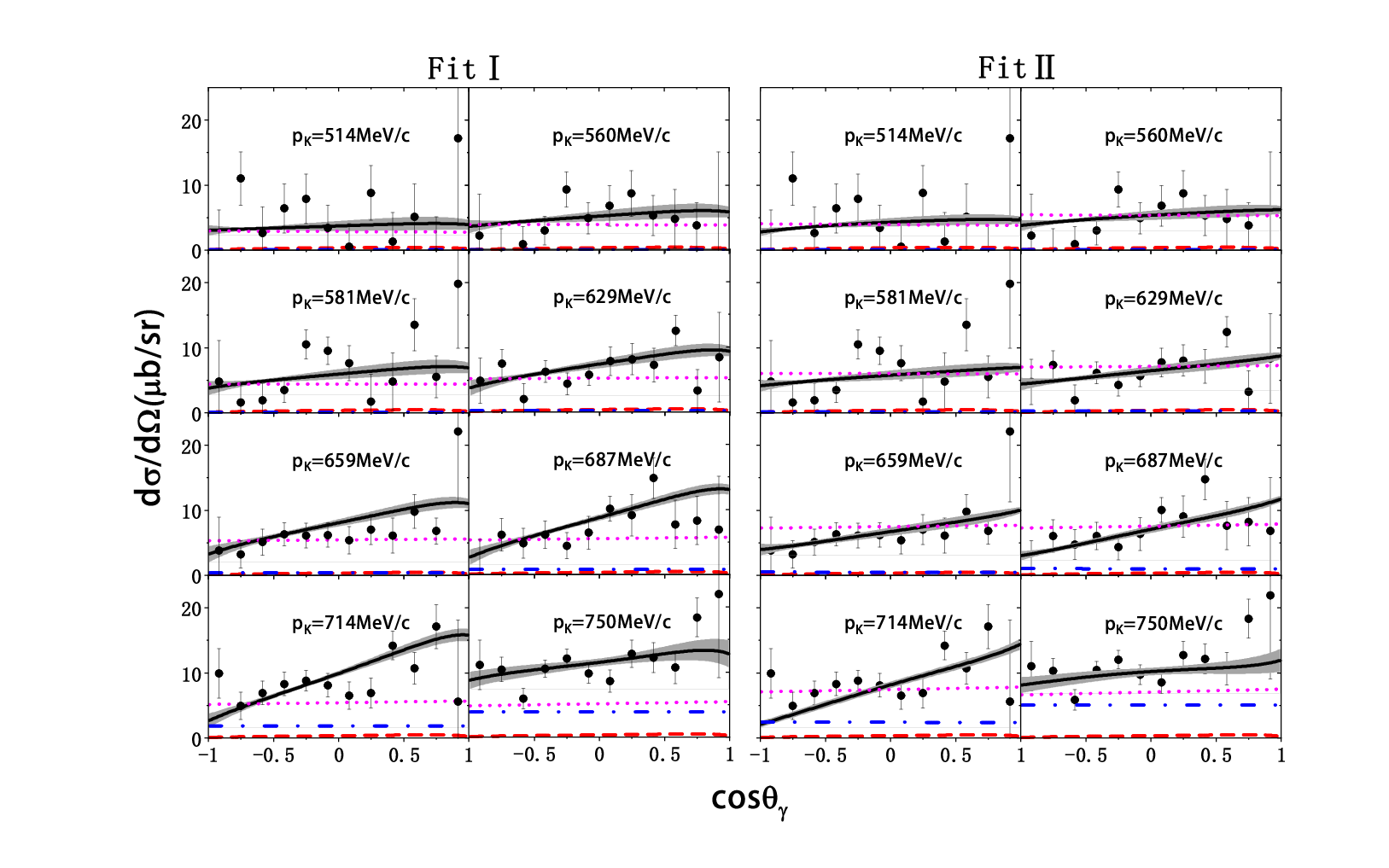} 
	\caption{Differential cross sections for the $K^- p \rightarrow \gamma \Sigma$ reaction as a function of $\cos\theta_\gamma$ in the center-of-mass frame in Model I. The solid, dashed, dotted, and dot-dashed lines represent the full, background, $\Lambda(1600)$, and $\Lambda(1670)$ contributions, respectively. The shaded bands correspond to the $1\sigma$ error regions of the fitting results.}
		\label{fig:model1_diff_xsec}
	\end{center}
\end{figure*}
Given that the knowledge of the radiative decays for these $\Lambda$ resonances is still very limited, we treat the products of the coupling constants, $g_{R\Sigma\gamma} g_{RNK}$, as free parameters to be determined by fitting to the experimental data. Furthermore, we introduce a phase factor($e^{i\phi_R}$) for each resonance amplitude, which is also fitted to the data. For the background contributions, thanks to extensive studies on the crossing-related reaction $\gamma p \to K^+ \Sigma^0$, we adopt the coupling constants, form factors, and gauge-invariance restoration method from previous works on the photoproduction channel. Consequently, there are no free parameters in the background terms considered in this work. With the help of the cernlib package MINUIT and the formalisms presented in the last section, the free parameters are fitted to the experimental data.

As a first step, following the approach in the experimental analysis of Ref.\cite{EXData1_Prakhov}, it is instructive to perform a fit considering only the background terms plus a single $\Lambda$ resonance. Figure~\ref{fig:xsec1R} shows the best-fit results for the total cross sections in this single-resonance scenario.
By inspecting Fig.~\ref{fig:xsec1R}, a comparison with the analysis of the $K^- p \to \gamma\Lambda$ reaction\cite{Pan:2025fzg} is warranted. Notably, the background contributions in the $\gamma\Sigma$ channel are significantly smaller than in the $\gamma\Lambda$ channel. This can be understood by the fact that most background terms scale with the $\Sigma NK$ coupling constant, and the $\Sigma NK$ coupling is substantially weaker than the $\Lambda NK$ coupling.

Although the single-resonance models do not provide a satisfactory description of the data, the results clearly reveal the crucial role of the $\Lambda(1600)$. Specifically, the data at lower energies cannot be reproduced without the inclusion of the $\Lambda(1600)$ resonance. Conversely, including only the $\Lambda(1600)$ leads to a poor description of the data at higher energies. This observation indicates that additional resonance contributions are necessary to improve the description of the data.

\begin{figure*}[htbp]
	\begin{center}
		\hspace*{-25mm}
		\includegraphics[scale=0.8]{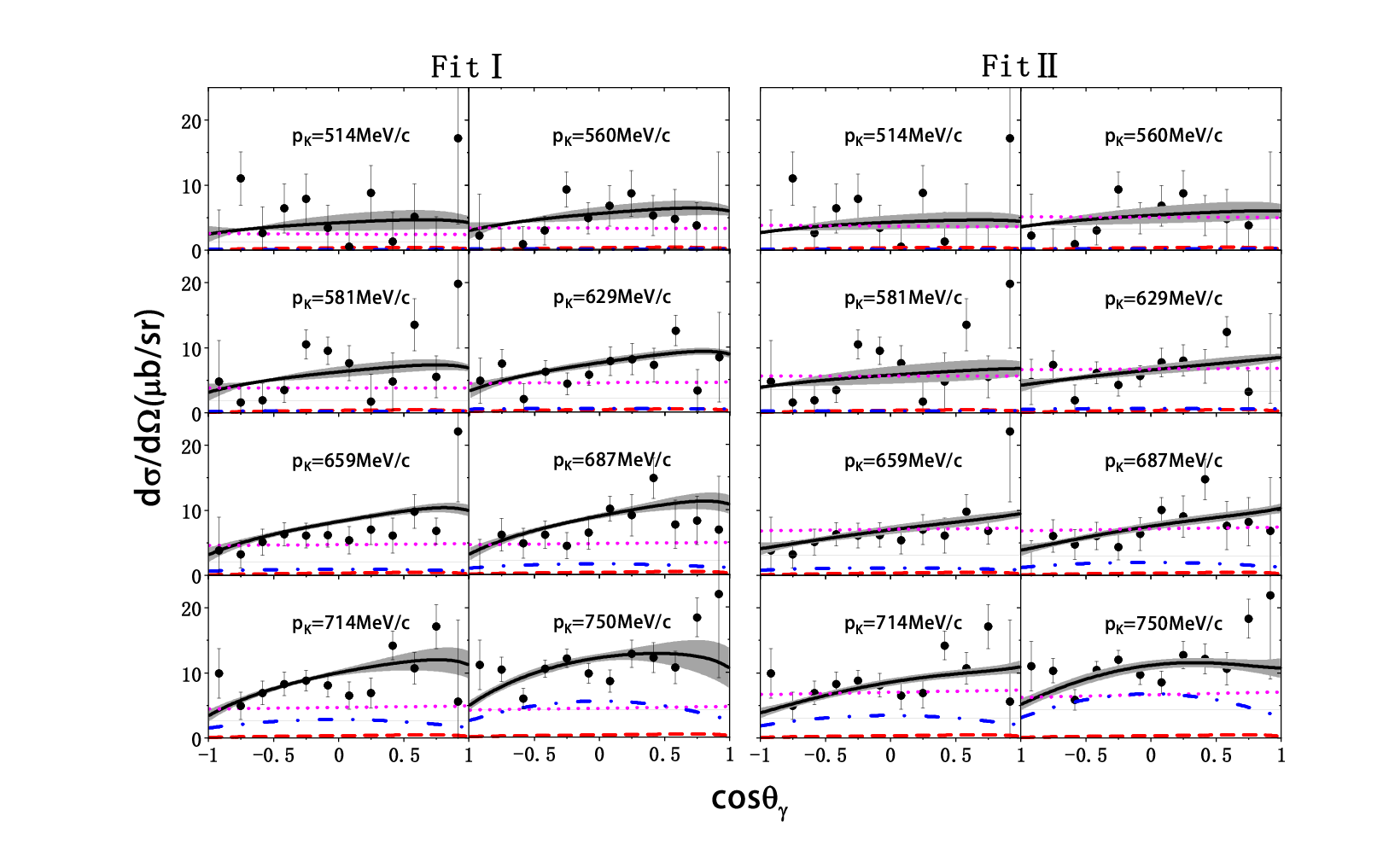} 
	\caption{Differential cross sections for the $K^- p \rightarrow \gamma \Sigma$ reaction as a function of $\cos\theta_\gamma$ in the center-of-mass frame in Model II. The solid, dashed, dotted, and dot-dashed lines represent the full, background, $\Lambda(1600)$, and $\Lambda(1690)$ contributions, respectively. The shaded bands correspond to the $1\sigma$ error regions of the fitting results.}
		\label{fig:model2_diff_xsec}
	\end{center}
\end{figure*}

\begin{figure}[htbp]
	\begin{center}
		\hspace*{-5mm}
		\includegraphics[scale=0.4]{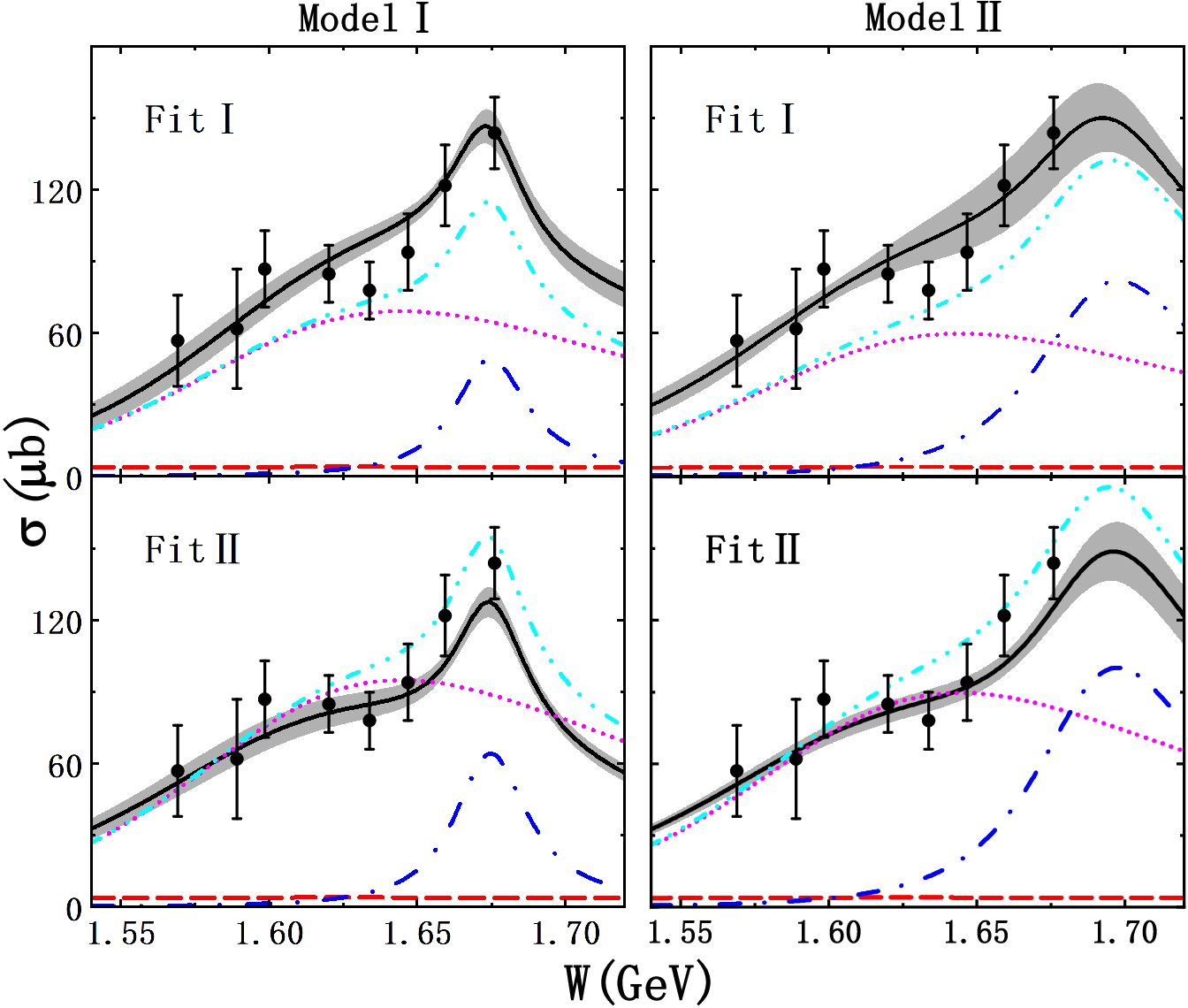} 
		\caption{Total cross sections for the $K^- +p \rightarrow \gamma +\Sigma$ reaction for Model I (Left) and Model II (Right). The shaded bands represent the $1\sigma$ error regions of the fitting results, reflecting uncertainties due to $1\sigma$ parameter variations. The solid, dotted, and dashed lines denote the contributions of the full model, the $\Lambda(1600)$, and the background terms, respectively. The dash-dotted line shows the contribution of either $\Lambda(1670)$ (in Model I) or $\Lambda(1690)$ (in Model II). The dash-dot-dotted line illustrates the corresponding sum of the resonance contributions.}
		\label{fig:total_xsec}
	\end{center}
\end{figure}

     \begin{table}[htbp]
        \caption{Fitted parameters for Model I.}
        \begin{tabular}{cccc}
            \hline\hline
            Parameters                                                              &  Fit I                                &Fit    II  \\
            \hline
$g_{\smsm{\Lambda(1600) \bar K N }}g_{\smsm{\Lambda(1600) \Sigma \gamma}} $ &$ 14.872\pm0.850 $                 &$ 17.430\pm0.920 $ \\
$\phi_{\Lambda(1600)} $                                                     &$ 0.569\pm0.641 $                      &$ -2.875\pm0.495 $     \\
$g_{\smsm{\Lambda(1670) \bar K N }}g_{\smsm{\Lambda(1670) \Sigma \gamma}} $ &$ -1.203\pm 0.123$                     &$ -1.3725\pm 0.094$\\
$\phi_{\Lambda(1670)} $                                                     &$ -0.540\pm0.704 $                     &$ 2.121\pm 0.457 $ \\

            \hline \hline
        \end{tabular}
        \label{tab1670}
    \end{table}
To enhance the description of the data, we first consider Model~I, which incorporates contributions from the background terms, the $\Lambda(1600)$, and the $\Lambda(1670)$. By fitting this model to the experimental data, we get two equally good fitting results, each exhibiting different interference patterns. The resulting fitted parameters are listed in Table~\ref{tab1670}, and the corresponding differential and total cross sections are shown in Figs.~\ref{fig:model1_diff_xsec} and \ref{fig:total_xsec}, respectively. In Fit I, the interference between the resonance contribution and the background is constructive, leading to relatively small individual contributions from the two resonances. Conversely, in Fit II, the interference is destructive, resulting in relatively large individual contributions from the two resonances.

Both Fit I and Fit II of Model~I provide a satisfactory description of the data, with  $\chi^2/\text{d.o.f.}$ values of $0.976$ and $0.97$, respectively.  Figs.~\ref{fig:model1_diff_xsec} and \ref{fig:total_xsec} show the background contribution (dashed line) to be minor. The $\Lambda(1600)$ (dotted line) dominates across the whole energy range under study. However, the $\Lambda(1670)$ resonance (dash-dotted line) also exhibits a significant contribution around its mass peak, as further illustrated in the left panel of Fig.~\ref{fig:total_xsec}.
Furthermore, significant interference effects between the resonant and background amplitudes are essential for describing the experimental data. This is evident from the full result (solid line), which deviates substantially from the sum of the individual contributions, and is crucial for achieving good agreement with the measurements.
    \begin{table}[htbp]
        \caption{Fitted parameters for Model II.}
        \begin{tabular}{cccc}
            \hline\hline
Parameters                                                              &  Fit I                                &Fit    II  \\
\hline
$g_{\smsm{\Lambda(1600) \bar K N }}g_{\smsm{\Lambda(1600) \Sigma \gamma}} $ &$ 13.830\pm0.618$                      &$ 16.924\pm1.140 $ \\
$\phi_{\Lambda(1600)} $                                                     &$ 1.249\pm1.536 $                      &$ -3.005\pm0.555$  \\
$g_{\smsm{\Lambda(1690) \bar K N }}g_{\smsm{\Lambda(1690) \Sigma \gamma}} $ &$ 17.612\pm 3.361$                     &$ 19.484\pm 1.230$\\
$\phi_{\Lambda(1690)} $                                                     &$ 2.065\pm1.646 $                      &$ -2.372\pm 0.465$ \\
\hline \hline

        \end{tabular}
        \label{tab1690}
    \end{table}

Alternatively, we also investigate Model~II, which incorporates the $\Lambda(1690)$ resonance in addition to the background and the $\Lambda(1600)$. Since $J^P $ quantum numbers of the $\Lambda(1690)$ are $\frac{3}{2}^-$, the interaction Lagrangian for the $\Lambda(1690)\Sigma\gamma$ vertex could, in principle, contain two independent terms. To reduce the number of free parameters, we retain only the $\mathcal{L}^{3/2^-}_{R\Sigma\gamma,(2)}$ term. We have verified that using the $\mathcal{L}^{3/2^-}_{R\Sigma\gamma,(1)}$ term instead does not significantly alter the results.

This simplification results in a model with four free parameters, which are fitted to the experimental data. The resulting parameters are listed in Table~\ref{tab1690}. Here, we again obtain two solutions, each exhibiting a distinct interference pattern. As the differential cross sections in Fig.~\ref{fig:model2_diff_xsec} demonstrate, both solutions of Model II also provide a good description of the data, achieving $\chi^2/\text{d.o.f.}$ values of $0.998$ and $0.978$, respectively.

The total cross sections obtained from Model~I (left panel) and Model~II (right panel) are compared in Fig.~\ref{fig:total_xsec}. As shown in the figure, in all cases the $\Lambda(1600)$ resonance (dotted line) dominates at lower energies, accompanied by significant interference effects between the resonant and background contributions. For Model II, however, the $\Lambda(1690)$ contribution becomes dominant near its mass peak.
A primary distinction between the Model I and II emerges in the high-energy trend of the total cross section. This is driven by the differing contributions of the $\Lambda(1670)$ in Model~I and the $\Lambda(1690)$ in Model~II. In contrast to Model~I, where the total cross section begins to decrease at the highest energies measured, the cross section predicted by Model~II continues to rise gradually, peaking at a higher energy. This behavior is a direct consequence of the mass difference between the $\Lambda(1670)$ and $\Lambda(1690)$. Unfortunately, the current experimental data do not extend to the energy region required to distinguish between these two scenarios. Therefore, future measurements of this reaction at higher energies would be highly valuable for discriminating between the two models.

Furthermore, given that the $\Lambda(1670)$$(\frac{1}{2}^-)$ and $\Lambda(1690)$$(\frac{3}{2}^-)$ possess different quantum numbers, the two models are expected to yield different predictions for the angular distributions. Since these two resonances contribute mainly at higher energies, the models provide similar results at lower energies. However, at $p_K = 750$~MeV, the angular distributions from Model~II already show clear curvature, which is indicative of higher partial-wave contributions. This suggests that more precise angular distribution data at energies around the $\Lambda(1690)$ mass peak could offer further evidence for the role of this resonance in the reaction.

\begin{figure}[htbp]
	\begin{center}
		\includegraphics[scale=0.37]{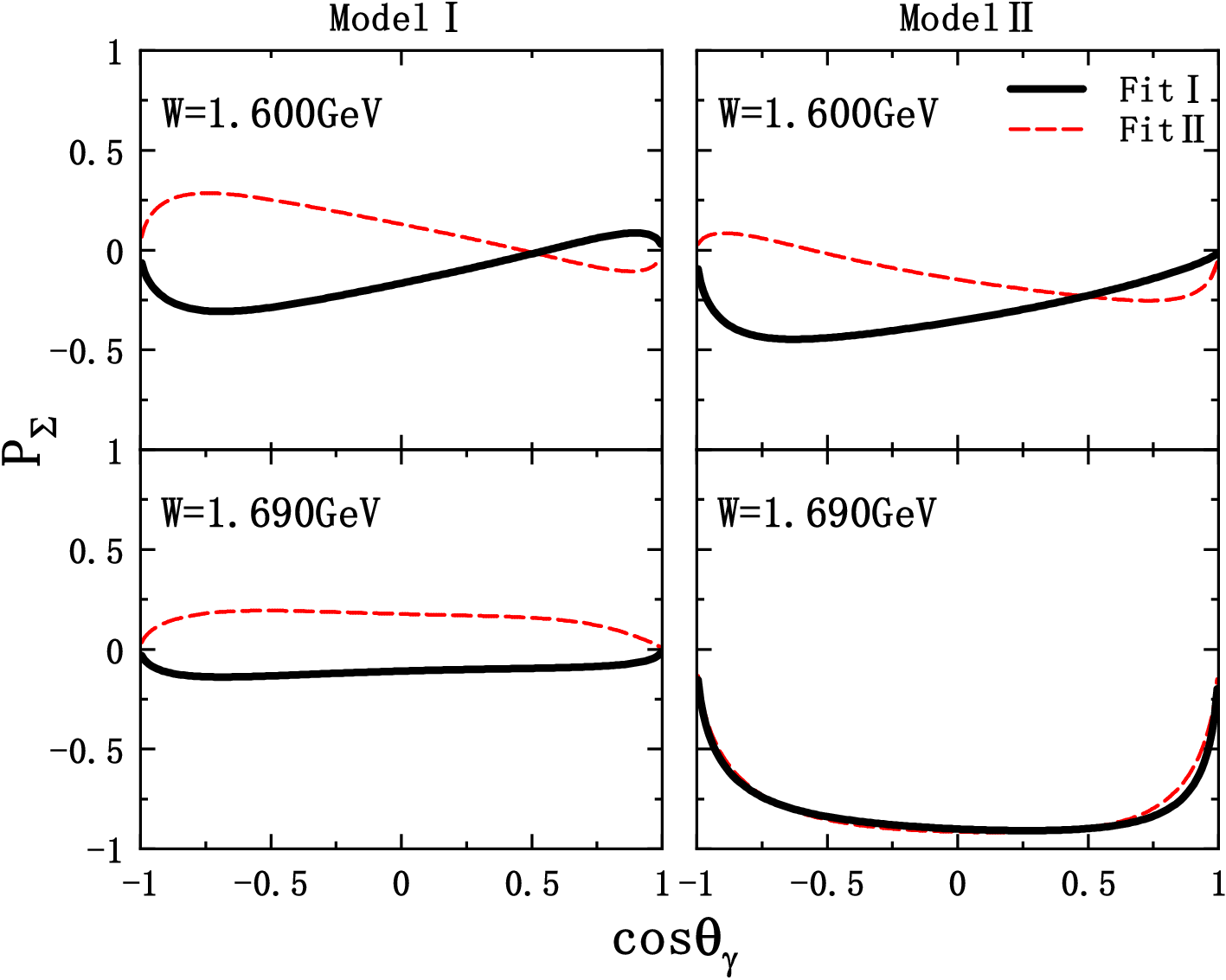} 
		\caption{Polarization of $\Sigma$ perpendicular to the scattering plane. The plot compares predictions from Model I (Left panel) and Model II (Right panel). Within each model, the solid and dashed lines correspond to the results of Fit I and Fit II, respectively.}
		\label{fig:polarization}
	\end{center}
\end{figure}

To further differentiate between the models and their respective solutions, we calculated the polarization of the final $\Sigma$. The results, presented in Fig.~\ref{fig:polarization}, compare the predictions of various solutions at two center-of-mass energies. Firstly, the predictions from Model I and Model II show distinct differences at W = 1.69 GeV, an energy where the $\Lambda(1690)$ contribute most significantly. Therefore, measurements of the $\Sigma$ polarization at W=1.69 GeV would be helpful in identifying the role of the $\Lambda(1690)$ in this reaction. On the other hand, at lower energy(W=1.6 GeV) the two models yield similar predictions. This is due to the minor roles played by $\Lambda(1670)$ and $\Lambda(1690)$ at this energy. Therefore, in both models, the $\Sigma$ polarization is primarily driven by the interference between the $\Lambda(1600)$ and background terms, which produce comparable results. To distinguish the two solutions within each model, as illustrated in Fig.\ref{fig:polarization}, measurements of $\Sigma$ polarization at lower energies are necessary. At these energies, the solutions exhibit distinct predictions at backward angles. It is also noteworthy that for Model II, the two solutions provide almost identical predictions for the $\Sigma$ polarization. This can be understood by considering that at W = 1.69 GeV, the $\Lambda(1600)$ and $\Lambda(1690)$ provide the most significant contributions in Model II, and the $\Sigma$ polarization arises from the interference between these two resonances. As shown in Tab.\ref{tab1690}, the relative phase between these two resonances is almost identical in both solutions, leading to the observed similarity in their $\Sigma$ polarization predictions. These calculated results demonstrate that the $\Sigma$ polarization is highly sensitive to the underlying resonance contributions. Consequently, it could serve as a powerful tool for differentiating between theoretical models in future experiments.

\section{SUMMARY}\label{sec:summary}

We have investigated the $K^- p \rightarrow \gamma \Sigma^0$ reaction using an effective Lagrangian approach, considering contributions from background terms and hyperon resonances. Our analysis shows that a single-resonance model is insufficient to describe the data. Consequently, we considered two alternative models: Model~I, which includes the $\Lambda(1600)$ and $\Lambda(1670)$, and Model~II, featuring the $\Lambda(1600)$ and $\Lambda(1690)$. For each model, we obtain two equally good fitting solutions. In all cases, we find the $\Lambda(1600)$ plays an essential role at lower energies, while the contribution from the higher-mass resonance becomes significant at higher energies. Furthermore, we also find that the interference between resonance and background amplitudes is crucial to reproduce the data.

Although both models fit the current data, they yield distinct predictions for total cross sections at higher energies and for the $\Sigma$ polarization. These observables are sensitive to the resonance content and thus offer a clear way to experimentally distinguish various theoretical models in the future.

\begin{acknowledgements}
We acknowledge the supports from the Natural Science Foundation of Shaanxi Province under Grants No.2024JC-YBMS-010.
\end{acknowledgements}


\end{document}